# Inverse designing metamaterials with programmable nonlinear functional responses in graph space

Marco Maurizi[1]*†, Derek Xu[2]†, Yu-Tong Wang[3]†, Desheng Yao[1], David Hahn[1], Mourad Oudich[3,4], Anish Satpati[1], Mathieu Bauchy[5], Wei Wang[2], Yizhou Sun[2], Yun Jing[3], Xiaoyu Rayne Zheng[1]*

[1]Department of Materials Science and Engineering, University of California Berkeley; Berkeley, 94720, USA.

[2]Department of Computer Science, University of California Los Angeles; Los Angeles, 90095, USA.

[3]Graduate Program in Acoustics, Penn State University; Pennsylvania 16802, USA.

[4]Universite' de Lorraine, CNRS, Institut Jean Lamour; F-54000 Nancy, France.

[5]Department of Civil and Environmental Engineering, University of California Los Angeles; Los Angeles, 90095, USA.

*Corresponding authors. Emails: marcomaurizi@berkeley.edu, rayne23@berkeley.edu

†These authors contributed equally to this work.

## Abstract

Material responses to static and dynamic stimuli, represented as nonlinear curves, are design targets for engineering functionalities like structural support, impact protection, and acoustic and photonic bandgaps. Three-dimensional metamaterials offer significant tunability due to their internal structure, yet existing methods struggle to capture their complex behavior-to-structure relationships. We present GraphMetaMat, a graph-based framework capable of designing three-dimensional metamaterials with programmable responses and arbitrary manufacturing constraints. Integrating graph networks, physics biases, reinforcement learning, and tree search, GraphMetaMat can target stress-strain curves spanning four orders of magnitude and complex behaviors, as well as viscoelastic transmission responses with varying attenuation gaps. GraphMetaMat can create cushioning materials for protective equipment and vibration-damping panels for electric vehicles, outperforming commercial materials, and enabling the automatic design of materials with on-demand functionalities.

## Introduction

Architecting the structure of metamaterials across scales allows for achieving properties and functionalities not found in classical engineering materials, significantly expanding the traditional design space (*1–6*). Examples include periodic truss metamaterials with high stiffness- and strength-to-weight ratios (*7*, *8*), lattice microstructures with tunable negative Poisson's ratio (*9*, *10*), and programmable anisotropic stiffness (*11*). Traditionally, the design process has been guided by engineering intuition (*12*), nature-inspiration (*13*) or topology optimization (*14*), hinging on prior expertise or computationally intensive calculations. While existing machine learning-driven approaches efficiently optimize specific properties (*10*, *15–23*), a key challenge remains in capturing complex, on-demand functionalities (*24*).

The characteristic fingerprints of materials, such as static and dynamic stress-strain responses, wave transmission and absorption responses to dynamic loadings, are influenced by the nonlinear interplay between constitutive material behavior, architecture, and applied loads. Metamaterials offer broad control over these responses (*25*, *26*), but this tunability comes with increased complexity. As the design space expands to include intricate structures and complex material behaviors, the data required explodes, making the process computationally or experimentally expensive and limiting design diversity (*25*, *27*). Most current design approaches for metamaterials rely on preassigned parametrization groups (*28*) or combinations of existing unit cell designs (*15*, *29*, *30*). Moreover, as most inverse design approaches depend on gradient-based algorithms (*25*, *27*, *31*), enforcing geometrical and manufacturing constraints becomes challenging, exacerbating the data-hungry issue.

To address these challenges, we introduce GraphMetaMat, a graph-based framework that uses graph neural networks (GNNs) as basic building blocks. It combines deep imitation learning (IL) and reinforcement learning (RL) with Monte Carlo tree search (MCTS) as an unsupervised generative model to inverse design graph-representable metamaterials with prescribed nonlinear functional responses (e.g., stress-strain and wave transmission curves) resulting from complex physical phenomena (e.g., buckling, contact, resonance, and damping). Our framework architecture can accommodate any graph-representable metamaterial and nonlinear physical response, incorporating arbitrary geometric constraints. We focus on the inverse design of truss metamaterials with prescribed quasi-static stress-strain and wave transmission curves as impactful proofs of concept.

Our results demonstrate that GraphMetaMat can rapidly generate metamaterial designs with stress responses spanning four orders of magnitude up to 30 % of strain and wave transmission curves with tunable attenuation gaps (i.e., low transmission values) in the frequency range 1 – 12 kHz. Additionally, applying our framework to real-world engineering problems, we show that GraphMetaMat can discover lightweight metamaterials with high energy absorption yet low peak stress for protective equipment and with low vibration transmission for noise-reduction panels in electric vehicles. While generalization to different loading conditions (e.g., multi-axial loading) and constitutive materials requires additional data, our framework provides a tool to navigate and unify the vast design space enabled by architected materials and additive manufacturing, setting the stage for fully automated discovery and design of metamaterials.

Results

Metamaterial-to-graph space

Translating metamaterials into graphs allows us to exploit the inductive biases of GNNs (*32*). In this 'graph space' (Fig. 1A), metamaterial struts and junctions (intersections between struts) are represented as graph edges (*E*) and nodes (*V*), respectively. The *geometry*, including node locations and strut shapes and dimensions (e.g., strut radius), is hence encoded into node ($v_i$) and edge ($e_{ij}$) features (see Materials and Methods). The *topology* of the metamaterial, i.e., which struts connect to each other, is captured by graph connectivity, eliminating the need to parametrize the design space with pre-existing structures (*28*, *30*, *33*) or pre-selected building blocks (*29*, *34*).

To test our inverse design framework, we restricted the design space to cubic symmetric periodic truss metamaterials (Fig. 1A and figs. S1 and S5), similarly to (*35*). While we adopted this design space, we emphasize that any three-dimensional (3D) truss architecture (either homogeneous or heterogeneous, Fig. 1, A and B, respectively), made of any constitutive material, can be represented using graphs, without limitations to regular periodic (*27*, *36*, *37*) or 2D structures (*29*, *38*, *39*). Cubic symmetry ensures invariant mechanical properties along three orthogonal directions and facilitates the generation of heterogeneous metamaterials by modifying only the interior nodes and connectivity (Fig. 1B), thus avoiding boundary node-matching methods (*34*). The chosen design space also includes classical stretching- and bending-dominated lattices, such as octet and Kelvin foams, as well as bi-stable structures (fig. S3). Detailed information is provided in Materials and Methods and Supplementary Text.

This potentially unlimited design space, while challenging, offers the opportunity to design metamaterials with a wide variety of physical responses. To train our framework, we collected ~ 3000 graph-curve data points via high-fidelity simulations for quasi-static stress-strain and wave transmission responses (see Materials and Methods). Each graph represents a metamaterial with varying geometry, topology, and relative density, $\bar{\rho}$ (ranging from 5 to 25 %). Figure 1, C and D show the corresponding response space of our datasets. The compressive and vibrational responses exhibit a rich diversity of complex behaviors, as highlighted by the three examples for each response type (see also figs. S11 and 20). This vast design space enables our framework to inverse design target functional responses, spanning orders of magnitude in stress and featuring complex transmission characteristics.

**GraphMetaMat – a framework for the design of metamaterials**

Our inverse-design framework, GraphMetaMat, is illustrated in Fig. 2. Initially, we train a forward GNN-based model on simulated data to learn the behavior of metamaterials in the graph space, mapping structure to functional response (Fig. 2C and fig. S23). This model is then used as a fast surrogate predictor within the inverse-design framework. Next, in an IL setting, a GNN-based policy network, $\pi_\theta$, is pre-trained to learn the correct sequence of actions, $\{\boldsymbol{a}_0, \ldots, \boldsymbol{a}_K\}$, including nodes, edges and relative density, $\bar{\rho}$ (Fig. 2A, upper left), needed to reconstruct the ground-truth graphs ($\overline{G}$) associated with the training input curves. Given the current state $\boldsymbol{s}_k$, composed of the current graph ($G_k$) and the target curve ($\bar{y}$), $\pi_\theta$ provides the action $\boldsymbol{a}_{k+1}$ to build the next graph state ($\boldsymbol{s}_{k+1}$). After IL, RL fine-tunes $\pi_\theta$ to generate graphs whose responses closely match the desired input curves (plot in Fig. 2C). The policy network's weights are updated using a proximal policy optimization scheme (*40*), aiming to maximize the future expected reward $R = w_J J - w_U U$, where $J$ measures the similarity between the curves, $U$ is the forward model's uncertainty, and $w_J$ and $w_U$ are weighting hyperparameters. This approach overcomes the typical one-to-many mapping issue (i.e., multiple designs for a given target response) by learning the conditional probability and sampling from it to get a unique solution.

Once trained, we use MCTS rollouts to estimate the value of each state $\boldsymbol{s}_k$ in a search tree (Fig. 2A, upper right) by sampling each action $\boldsymbol{a}_{k+1} \sim \pi_\theta(\cdot | \boldsymbol{s}_k)$ from its probability distribution over actions. The model iteratively samples multiple generated graphs and selects the best one from 128 iterations, significantly improving inverse design performance (fig. S35). As a validation step, the corresponding structure is 3D-printed and tested (Fig. 2D). Complete technical details are provided in Materials and Methods and Supplementary Text.

Overall, by employing IL for pre-training on graph-labeled training curves, RL for fine-tuning on unlabeled curves, and MCTS for refining the search method, GraphMetaMat can generate a graph corresponding to a metamaterial with the desired behavior (within an average tolerance of 10%) and adhere to any geometric constraints (e.g., manufacturing constraints).

*Inverse design of user-defined responses*

Designing structures to match unseen target curves with known associated graphs (i.e., test set) is a meaningful task to assess model performance (fig. S28). However, the inverse design of user-defined curves, where the existence of an associated structure is not guaranteed, is more challenging. To address this, we designed four stress-strain and three wave transmission response types that extend beyond the dataset response space (Fig. 3, A and D). Key to this task were the physics bias for stress predictions (fig. S23, bottom diagram), extended $\bar{\rho}$ search (2 – 30 %), and pre-training of the forward model (~ 90 % accuracy on 3000 labeled data) (fig. S25).

The four classes of stress-strain responses are: (i) stiffer and stronger than the most rigid structure in the dataset; (ii) convex with strain hardening; (iii) concave with strain hardening and/or softening; and (iv) softer than the most compliant structure in the dataset (see Materials and Methods). For each type, Fig. 3B shows a randomly selected target response, the corresponding prediction, the simulation of the generated graph (Fig. 3C), and the closest curve from the training dataset (best train match). The relative error between the target and predicted curves ranges from 1.98 to 6.85 %. While simulations of the generated structures exhibit higher errors (from 1.8 to 30 %), the qualitative behavior is remarkably captured, even outside the training range of $\bar{\rho}$. Generalizing on stiffer and stronger structures (i) appears more challenging than for more compliant structures (iv), likely due to the higher proportion of softer structures in the training set (fig. S15, A and B). When benchmarked against the best train match, GraphMetaMat shows similar or larger errors for types (i), (ii), and (iii), except for (iv), where softer designs seem to be more easily accessible. Nonetheless, comparing error distributions on 2,000 targets, GraphMetaMat outperforms in average both the best train match (fig. S33A) and random structure selection (fig. S35B, no search and untrained policy network) by ~ 3 and 263 times, respectively.

Defining realistic target transmission curves is more challenging than shaping realistic stress-strain responses. Applications often require metamaterials with specific attenuation gaps (low transmission in certain frequency ranges) rather than exact transmission values. To this end, instead of targeting transmission curves $T(f)$, we target binary sequences defined by thresholding $T$ with $T_{th}$, where '0' corresponds to 'low transmission' ($T < T_{th}$) and '1' to 'high transmission' ($T \geq T_{th}$). During training, GraphMetaMat still predicts the transmission curve, which is then binarized for comparison with the target sequence. To challenge the model on the inverse design of tunable attenuation gaps ($T < T_{th}$), we defined three target types with varying gap sizes $\Delta f$, from ~1.4 to 2.1 and 2.7 kHz, respectively (Fig. 3D). Various sequences were generated by shifting the central frequency of the two gaps (Materials and Methods). We set $T_{th} = -10\ dB$, a reasonable threshold for vibration attenuation. Figure 3E shows one example of inverse design per type. GraphMetaMat's predictions and simulations of the generated structures are benchmarked against the best train matches. GraphMetaMat exhibits an accuracy on the target attenuation gaps of ~ 0.8, outperforming the best train match by ~108, 58, and 17 % for types (i), (ii), and (iii), respectively. The corresponding simulated responses show lower accuracy due to the forward model's error but still outperform the best train matches by ~ 30, 32, and 6 %. Unlike stress-strain predictions, the accuracy distribution of model predictions on target sequences (fig. S33C) is only comparable to that of the best train match. Yet, GraphMetaMat outperforms random structure selection (no search and untrained policy network) by ~ 1.3 times on average (fig. S35, E).

**Energy-absorbing and noise-control applications**

To challenge our model, we applied GraphMetaMat to design cushioning material for a lacrosse chest protector and a vibration-damping panel for electric vehicles. Figure 4A shows a commercial lacrosse chest protector composed of four different foam layers (inset in Fig. 4B and fig. S37, A and B). This foam sandwich is designed to prevent *commotio cordis*, "the second leading cause of death among young athletes" (*41*), which occurs upon a 30 – 50 mph impact to the chest, causing ventricular fibrillation (*42*). To mitigate this phenomenon, standards (*43*) require specific transmitted peak forces for certified protectors. Although designing a certified protector under dynamic impact is beyond the scope of this work, we aimed to design 3D-printed cellular metamaterials with energy-absorbing performance comparable or superior to that of commercial protector foams. Specifically, reducing the transmitted peak force translates to simultaneously lowering the peak stress ($\sigma_{max}$) and increasing the energy absorption ($U$) (area underneath the curve) during impact (Supplementary Text "Energy-absorbing and noise-control applications"). To achieve this goal, using the measured average compressive response of the foam sandwich as a baseline (Fig. 4B), we targeted stress-strain curves with lower peak stress and the same or higher energy absorption (Fig. 4C) (Materials and Methods). To reduce the cost of high strain-rate training data, we first targeted quasi-static responses with our model and verified the results with dynamic impact simulations (fig. S38, F to H).

Restricting the design to structures with relative density between 5 and 10 %, Fig. 4D shows an example of metamaterial generated to achieve a response with 30 % lower peak stress and 20 % higher energy absorption. While the predicted response of the generated structure closely matches the target curve with an average relative error of ∼ 4.3 % (fig. S35C), the corresponding experimental error is higher at ∼ 27 %. Despite this, the target shape is well captured (inset up to 30 % of strain in Fig. 4D), and the inverse-designed metamaterial outperforms the commercial four-layer sandwich foams by exhibiting similar peak stress (∼ 0.003 MPa) yet ∼ 53 % higher energy absorption (Fig. 4E). Similar performance improvements over classical 3D-printed Kelvin foam and octet architectures are also shown in Fig. 4E. Owing to its periodicity, the generated metamaterial demonstrates a prolonged stress plateau compared to the protector's foams, thereby achieving higher energy absorption efficiency (fig. S38A). Although Kelvin foam and octet structures exhibit similar longer stress plateau, the generated design combines a higher stiffness with buckling (inset and snapshots in Fig. 4D) to achieve the target requirements. Simulations assuming the same constitutive material (fig. S38) confirm that our results primarily stem from geometrical and topological effects, and hold under both quasi-static and high strain-rate impact conditions (strain rate, $\dot{\varepsilon} \sim 344 \ s^{-1}$).

Fig. 5A shows the various sources of noise inside an electric vehicle (*44*). Although electric motors are generally quieter than internal combustion engines, the pure tonal noise at frequencies above 1 kHz (Fig. 5B) generated by electromagnetic forces is perceived as more annoying (*45*). This tonal noise varies with speed, with frequency increasing at higher speeds. Focusing on structure-borne noise, generated by structural vibrations in the frequency range 1 – 12 kHz (Fig. 5, A and B), we aimed to design metamaterials with broadband low transmission response. To achieve this and to challenge our model on tunable transmission levels, we considered constant target curves from 5 down to −40 dB (Fig. 5C). Figure 5E shows four examples of inverse-designed structures for decreasing vibration transmission levels, with corresponding predictions, experiments, and benchmarks. The benchmark (*46*) was selected from state-of-the-art metamaterials optimized for broadband vibration filtering (*46*, *47*) (figs. S36). For each level, the generated structures display oscillating behaviors as expected, with average experimental values deviating by ∼100, 149, 16, 8 % from the targets for (i), (ii), (iii), (iv), respectively. As the transmission level is lowered, the predicted response shifts down accordingly, to $-40$ dB, corresponding to a structure with broadband low transmission ($T < -20$ dB). Interestingly, the relative density $\bar{\rho}$ of the inverse-designed structures gradually decreases from (i) to (iv) (Fig. 5D). A correlation between low transmission and $\bar{\rho}$ seems to exist, as observed in our dataset (figs. S21 and S22). This can be explained by the shift of the first resonance frequency ($f_1$) of the structure to lower frequencies, relying on simple scaling laws (Supplementary Text "Energy-absorbing and noise-control applications"). Nevertheless, the design of metamaterials with tunable transmission is enabled by the simultaneous control of topology and relative density (fig. S19). Unlike the benchmark, which has high transmission at frequencies below 4 kHz and achieves attenuation gaps with additional mass, our designs (iii) and (iv) mitigate structure-borne noise across the entire frequency range while remaining lightweight (Fig. 5D and fig. S36).

**Discussion**

Designing metamaterials with programmable physical responses addresses the need for customizable materials. In this work, we introduced GraphMetaMat, a graph-based framework for inverse designing metamaterials with diverse complex functional responses. Although tested on mechanical truss metamaterials, the framework can be extended to any graph-representable metamaterials and physical responses. Compared to graph-based generative models, such as generative adversarial networks (*48*) and variational autoencoders (*49*), and prior inverse design frameworks (*27*, *29*, *31*, *50*), GraphMetaMat offers several advantages: (1) it explicitly imposes structural constraints (e.g., periodic repeatability, self-connectivity, manufacturability) via the action space, ensuring valid designs; (2) it handles any graph-based metamaterial, regardless of topology and size; (3) it operates at inference time generating hundreds of structures with different target responses without costly optimizations; (4) it controls both topology and relative density, resulting in a large design-response space. For simplicity, we focused on designing single functional responses. However, using the same constitutive materials, GraphMetaMat could be extended to design multiple responses simultaneously (e.g., responses along multiple directions), enabling rapid automatic design of multifunctional metamaterials.

**References and Notes**

1. H. Cui, D. Yao, R. Hensleigh, H. Lu, A. Calderon, Z. Xu, S. Davaria, Z. Wang, P. Mercier, P. Tarazaga, X. (Rayne) Zheng, Design and printing of proprioceptive three-dimensional architected robotic metamaterials. *Science* **376**, 1287–1293 (2022).

2. J. Bauer, L. R. Meza, T. A. Schaedler, R. Schwaiger, X. Zheng, L. Valdevit, Nanolattices: An Emerging Class of Mechanical Metamaterials. *Advanced Materials* **29**, 1701850 (2017).

3. E. T. Filipov, T. Tachi, G. H. Paulino, Origami tubes assembled into stiff, yet reconfigurable structures and metamaterials. *Proc. Natl. Acad. Sci. U.S.A.* **112**, 12321–12326 (2015).

4. C. M. Portela, B. W. Edwards, D. Veysset, Y. Sun, K. A. Nelson, D. M. Kochmann, J. R. Greer, Supersonic impact resilience of nanoarchitected carbon. *Nat. Mater.* **20**, 1491–1497 (2021).

5. T. A. Schaedler, A. J. Jacobsen, A. Torrents, A. E. Sorensen, J. Lian, J. R. Greer, L. Valdevit, W. B. Carter, Ultralight Metallic Microlattices. *Science* **334**, 962–965 (2011).

6. X. Xia, C. M. Spadaccini, J. R. Greer, Responsive materials architected in space and time. *Nat Rev Mater* **7**, 683–701 (2022).

7. X. Zheng, H. Lee, T. H. Weisgraber, M. Shusteff, J. DeOtte, E. B. Duoss, J. D. Kuntz, M. M. Biener, Q. Ge, J. A. Jackson, S. O. Kucheyev, N. X. Fang, C. M. Spadaccini, Ultralight, ultrastiff mechanical metamaterials. *Science* **344**, 1373–1377 (2014).

8. L. R. Meza, S. Das, J. R. Greer, Strong, lightweight, and recoverable three-dimensional ceramic nanolattices. *Science* **345**, 1322–1326 (2014).

9. S. Babaee, J. Shim, J. C. Weaver, E. R. Chen, N. Patel, K. Bertoldi, 3D Soft Metamaterials with Negative Poisson's Ratio. *Advanced Materials* **25**, 5044–5049 (2013).

10. T. Meier, R. Li, S. Mavrikos, B. Blankenship, Z. Vangelatos, M. E. Yildizdag, C. P. Grigoropoulos, Obtaining auxetic and isotropic metamaterials in counterintuitive design spaces: an automated optimization approach and experimental characterization. *npj Comput Mater* **10**, 3 (2024).

11. S. Xu, J. Shen, S. Zhou, X. Huang, Y. M. Xie, Design of lattice structures with controlled anisotropy. *Materials & Design* **93**, 443–447 (2016).

12. S. S. Injeti, C. Daraio, K. Bhattacharya, Metamaterials with engineered failure load and stiffness. *Proc. Natl. Acad. Sci. U.S.A.* **116**, 23960–23965 (2019).

13. M. C. Fernandes, J. Aizenberg, J. C. Weaver, K. Bertoldi, Mechanically robust lattices inspired by deep-sea glass sponges. *Nat. Mater.* **20**, 237–241 (2021).

14. S. Wu, Z. Luo, Z. Li, S. Liu, L.-C. Zhang, Topological design of pentamode metamaterials with additive manufacturing. *Computer Methods in Applied Mechanics and Engineering* **377**, 113708 (2021).

15. J.-H. Bastek, S. Kumar, B. Telgen, R. N. Glaesener, D. M. Kochmann, Inverting the structure–property map of truss metamaterials by deep learning. *Proc. Natl. Acad. Sci. U.S.A.* **119**, e2111505119 (2022).

16. S. Kumar, S. Tan, L. Zheng, D. M. Kochmann, Inverse-designed spinodoid metamaterials. *npj Comput Mater* **6**, 73 (2020).

17. B. Li, B. Deng, W. Shou, T.-H. Oh, Y. Hu, Y. Luo, L. Shi, W. Matusik, Computational discovery of microstructured composites with optimal stiffness-toughness trade-offs. *Sci. Adv.* **10**, eadk4284 (2024).

18. H. Liu, L. Li, Z. Wei, M. M. Smedskjaer, X. R. Zheng, M. Bauchy, De Novo Atomistic Discovery of Disordered Mechanical Metamaterials by Machine Learning. *Advanced Science* **11**, 2304834 (2024).

19. S. Luan, E. Chen, J. John, S. Gaitanaros, A data-driven framework for structure-property correlation in ordered and disordered cellular metamaterials. *Sci. Adv.* **9**, eadi1453 (2023).

20. Y. Mao, Q. He, X. Zhao, Designing complex architectured materials with generative adversarial networks. *Sci. Adv.* **6**, eaaz4169 (2020).

21. S. C. Shen, M. J. Buehler, Nature-inspired architected materials using unsupervised deep learning. *Commun Eng* **1**, 37 (2022).

22. S. Van 'T Sant, P. Thakolkaran, J. Martínez, S. Kumar, Inverse-designed growth-based cellular metamaterials. *Mechanics of Materials* **182**, 104668 (2023).

23. L. Zheng, S. Kumar, D. M. Kochmann, Data-driven topology optimization of spinodoid metamaterials with seamlessly tunable anisotropy. *Computer Methods in Applied Mechanics and Engineering* **383**, 113894 (2021).

24. X. Zheng, X. Zhang, T. Chen, I. Watanabe, Deep Learning in Mechanical Metamaterials: From Prediction and Generation to Inverse Design. *Advanced Materials* **35**, 2302530 (2023).

25. J.-H. Bastek, D. M. Kochmann, Inverse design of nonlinear mechanical metamaterials via video denoising diffusion models. *Nat Mach Intell* **5**, 1466–1475 (2023).

26. N. J. Gerard, M. Oudich, Z. Xu, D. Yao, H. Cui, C. J. Naify, A. Ikei, C. A. Rohde, X. (Rayne) Zheng, Y. Jing, Three-Dimensional Trampolinelike Behavior in an Ultralight Elastic Metamaterial. *Phys. Rev. Applied* **16**, 024015 (2021).

27. L. Zheng, K. Karapiperis, S. Kumar, D. M. Kochmann, Unifying the design space and optimizing linear and nonlinear truss metamaterials by generative modeling. *Nat Commun* **14**, 7563 (2023).

28. C. S. Ha, D. Yao, Z. Xu, C. Liu, H. Liu, D. Elkins, M. Kile, V. Deshpande, Z. Kong, M. Bauchy, X. Zheng, Rapid inverse design of metamaterials based on prescribed mechanical behavior through machine learning. *Nat Commun* **14**, 5765 (2023).

29. M. Maurizi, C. Gao, F. Berto, Inverse design of truss lattice materials with superior buckling resistance. *npj Comput Mater* **8**, 247 (2022).

30. A. J. Lew, K. Jin, M. J. Buehler, Designing architected materials for mechanical compression via simulation, deep learning, and experimentation. *npj Comput Mater* **9**, 80 (2023).

31. P. Thakolkaran, M. A. Espinal, S. Dhulipala, S. Kumar, C. M. Portela, Experiment-informed finite-strain inverse design of spinodal metamaterials. arXiv [Preprint] (2023). https://doi.org/10.48550/ARXIV.2312.11648.

32. K. Xu, W. Hu, J. Leskovec, S. Jegelka, How Powerful are Graph Neural Networks? arXiv [Preprint] (2018). https://doi.org/10.48550/ARXIV.1810.00826.

33. A. J. Lew, M. J. Buehler, Single-shot forward and inverse hierarchical architected materials design for nonlinear mechanical properties using an Attention-Diffusion model. *Materials Today* **64**, 10–20 (2023).

34. K. Liu, R. Sun, C. Daraio, Growth rules for irregular architected materials with programmable properties. *Science* **377**, 975–981 (2022).

35. J. Panetta, Q. Zhou, L. Malomo, N. Pietroni, P. Cignoni, D. Zorin, Elastic textures for additive fabrication. *ACM Trans. Graph.* **34**, 1–12 (2015).

36. P. P. Indurkar, S. Karlapati, A. J. D. Shaikeea, V. S. Deshpande, Predicting deformation mechanisms in architected metamaterials using GNN. doi: 10.48550/ARXIV.2202.09427 (2022).

37. L. Makatura, B. Wang, Y.-L. Chen, B. Deng, C. Wojtan, B. Bickel, W. Matusik, Procedural Metamaterials: A Unified Procedural Graph for Metamaterial Design. *ACM Trans. Graph.* **42**, 1–19 (2023).

38. M. Maurizi, C. Gao, F. Berto, Predicting stress, strain and deformation fields in materials and structures with graph neural networks. *Sci Rep* **12**, 21834 (2022).

39. D. Dold, D. Aranguren Van Egmond, Differentiable graph-structured models for inverse design of lattice materials. *Cell Reports Physical Science* **4**, 101586 (2023).

40. J. Schulman, F. Wolski, P. Dhariwal, A. Radford, O. Klimov, Proximal Policy Optimization Algorithms. doi: 10.48550/ARXIV.1707.06347 (2017).

41. E. I. Drewniak, D. B. Spenciner, J. J. Crisco, Mechanical Properties of Chest Protectors and the Likelihood of Ventricular Fibrillation Due to Commotio Cordis. *Journal of Applied Biomechanics* **23**, 282–288 (2007).

42. K. Kumar, S. N. Mandleywala, M. P. Gannon, N. A. M. Estes, J. Weinstock, M. S. Link, Development of a Chest Wall Protector Effective in Preventing Sudden Cardiac Death by Chest Wall Impact (Commotio Cordis). *Clinical Journal of Sport Medicine* **27**, 26–30 (2017).

43. N. Dau, C. Bir, E. McCalley, D. Halstead, M. S. Link, Development of the NOCSAE Standard to Reduce the Risk of Commotio Cordis. *Circ: Arrhythmia and Electrophysiology* **17** (2024).

44. A. K. Sahu, A. Emadi, B. Bilgin, Noise and Vibration in Switched Reluctance Motors: A Review on Structural Materials, Vibration Dampers, Acoustic Impedance, and Noise Masking Methods. *IEEE Access* **11**, 27702–27718 (2023).

45. B. Meek, H. Van Der Auwear, K. De Langhe, "Challenges in NVH for Electric Vehicles" in *Proceedings of the FISITA 2012 World Automotive Congress*, SAE-China, FISITA, Eds. (Springer Berlin Heidelberg, Berlin, Heidelberg, 2013; https://link.springer.com/10.1007/978-3-642-33777-2_56)vol. 191 of *Lecture Notes in Electrical Engineering*, pp. 675–685.

46. Muhammad, C. W. Lim, Phononic metastructures with ultrawide low frequency three-dimensional bandgaps as broadband low frequency filter. *Sci Rep* **11**, 7137 (2021).

47. F. Lucklum, M. J. Vellekoop, Bandgap engineering of three-dimensional phononic crystals in a simple cubic lattice. *Applied Physics Letters* **113**, 201902 (2018).

48. H. Wang, J. Wang, J. Wang, M. Zhao, W. Zhang, F. Zhang, X. Xie, M. Guo, GraphGAN: Graph Representation Learning With Generative Adversarial Nets. *AAAI* **32** (2018).

49. Q. Liu, M. Allamanis, M. Brockschmidt, A. L. Gaunt, Constrained Graph Variational Autoencoders for Molecule Design. arXiv [Preprint] (2018). https://doi.org/10.48550/ARXIV.1805.09076.

50. B. Deng, A. Zareei, X. Ding, J. C. Weaver, C. H. Rycroft, K. Bertoldi, Inverse Design of Mechanical Metamaterials with Target Nonlinear Response via a Neural Accelerated Evolution Strategy. *Advanced Materials* **34**, 2206238 (2022).

51. V. Dorléans, R. Delille, D. Notta-Cuvier, F. Lauro, E. Michau, Time-temperature superposition in viscoelasticity and viscoplasticity for thermoplastics. *Polymer Testing* **101**, 107287 (2021).

**Acknowledgments:** We acknowledge Olivia Chen for her assistance in characterizing the commercial chest protector foams. We also thank Haotian Lu for his help in setting up the vibration experiments. Additionally, we used OpenAI ChatGPT to proofread the abstract and main text of the manuscript.

    **Funding:**

        US National Science Foundation (NSF) DMREF grant 2119643

        US National Science Foundation (NSF) grant 1829071

        US National Science Foundation (NSF) grant 2106859

        US National Science Foundation (NSF) grant 2312501

        US National Science Foundation (NSF) grant 2312501

        US Defense Advanced Research Projects Agency (DARPA) grant HR00112490370

        US National Science Foundation (NSF) grant 2312501

        US National Science Foundation (NSF) DMREF grant 2119545

**Author contributions:**

    Conceptualization: XRZ, MM, WW, YS, MB, YJ,

    Methodology: MM, DX, YTW, DY, XRZ, WW, YS, MB, YJ

    Investigation: MM, DX, YTW, DY, DH, MO, AS

    Visualization: MM, YTW, DY

    Funding acquisition: XRZ, WW, YS, MB, YJ

    Project administration: XRZ, WW, YS, MB, YJ

    Supervision: XRZ, WW, YS, MB, YJ

    Writing – original draft: MM, DX

    Writing – review & editing: MM, XRZ, DX, YTW, DY, MO, WW, YS, YJ

**Competing interests:** Authors declare that they have no competing interests.

**Data and materials availability:** The datasets and codes developed in the current study will be freely open sourced at the time of publication.

# Figures

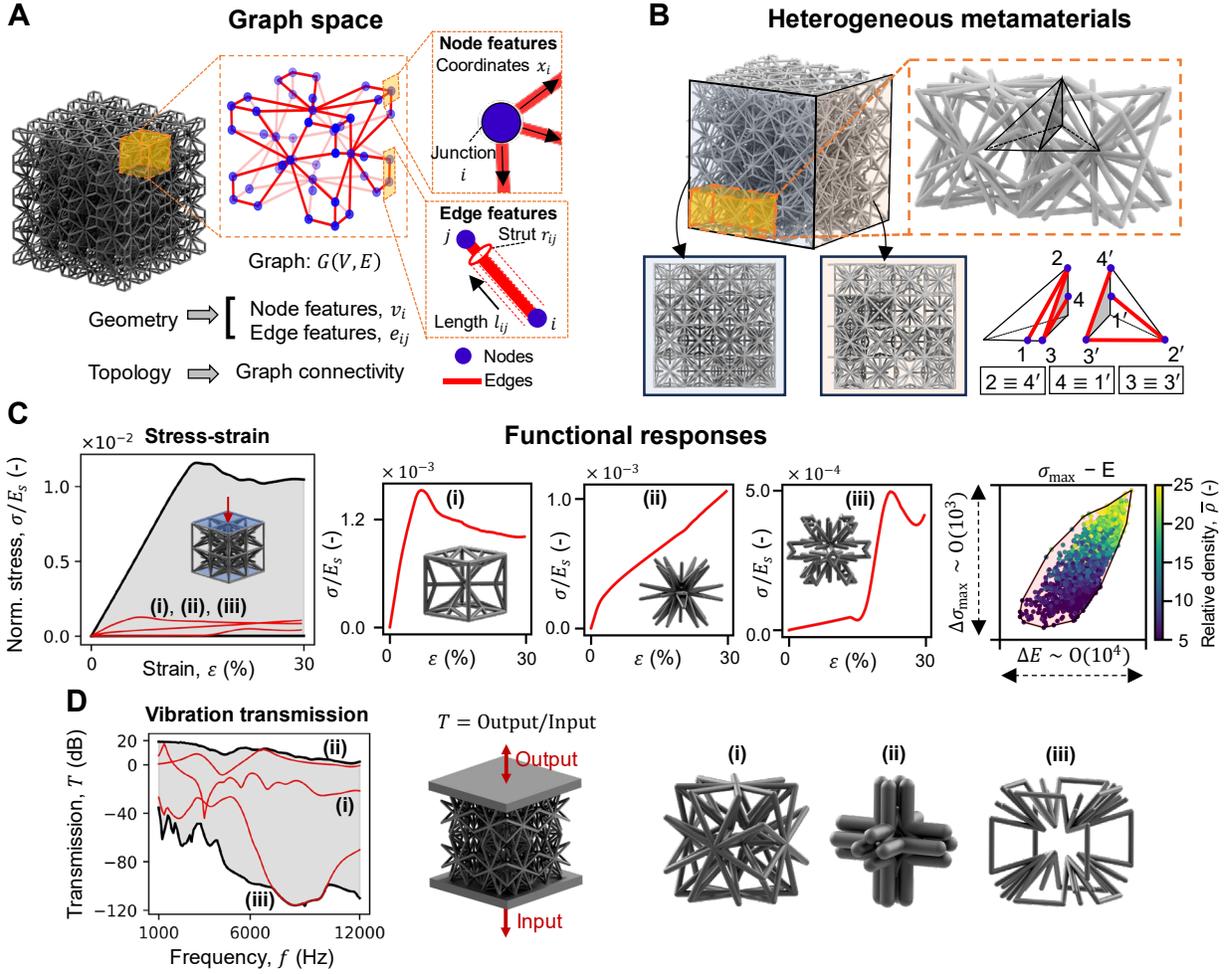

**Fig. 1. Graph-response design space.** (**A**) Graph space for metamaterials with geometrical and topological features. A periodic truss metamaterial is shown, together with the corresponding node and edge features used in the GNNs. The unit cell is translated onto a graph $G(V, E)$, with $V$ a collection of nodes connected through edges $E$. Geometric information is encoded into node $v_i$ and edge $e_{ij}$ features. Topological information is encoded as inductive bias into the graph connectivity. (**B**) Construction of heterogeneous metamaterials from individual unit cells belonging to our test design space (fig. S5). The graphs in the tetrahedra show the only needed constraints at the boundaries of the unit cells. (**C**) Stress-strain response space from the collected dataset. The stress is normalized by the constitutive material's Young's modulus $E_s$. The curves (i), (ii), and (iii) are three examples of macroscopic softening and hardening. The last plot on the right shows the corresponding strength ($\sigma_{max}$) – stiffness ($E$) property space, highlighting the orders of magnitude $O(10^m)$ spanned, with $m = 3$ for strength, and 4 for stiffness. (**D**) Vibration transmission response space from the collected dataset. (i), (ii), and (iii) identify responses with small attenuation gap, high, and low transmission, respectively. The sketch next to the response space shows how the transmission response is determined. It is measured as the log-ratio between the Fourier transforms of the output acceleration at the center point of the top plate and the input acceleration at the bottom plate (see also fig. S9).

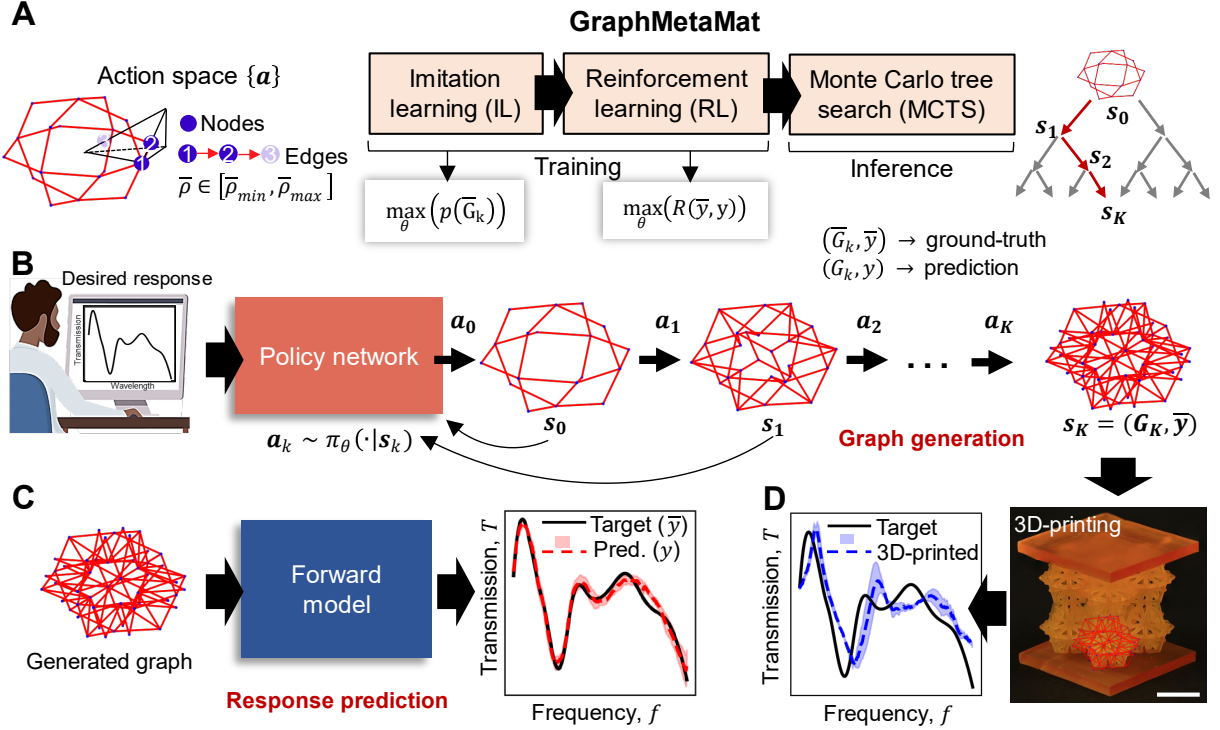

**Fig. 2. Overview of GraphMetaMat**. Example of inverse design of a metamaterial for target transmission response. The target curve is from the 90:5:5 split test set. (**A**) Steps of inverse design in graph space using IL, RL, and MCTS. The action space, $\{a\}$ is composed of all the possible nodes that can be selected, the edges that connect the chosen node with the previous node at each step $k$, and the relative density, $\bar{\rho}$. IL maximizes the probability $p(\bar{G}_k)$ of generating the true graph $\bar{G}_k$, associated with the target curve $\bar{y}$. RL maximizes the reward $R(\bar{y}, y)$, function of the error between the target ($\bar{y}$) and predicted ($y$) response. A schematic representation of the search tree executed during MCTS is illustrated on the right side. (**B**) A state-action ($s_k, a_k$) sequence predicted by the policy network $\pi_\theta(\cdot | s_k)$ for a given input desired response. (**C**) After each training iteration, the trained forward model is called to predict the functional response of the generated metamaterial. The error between this prediction ($y$) and the target response ($\bar{y}$) guides the training of the RL step. (**D**) 3D-printing and testing of the corresponding finite-size structure for validation. The plot compares the experimental transmission ('3D-printed') to the target response. Scale bar, 10 mm.

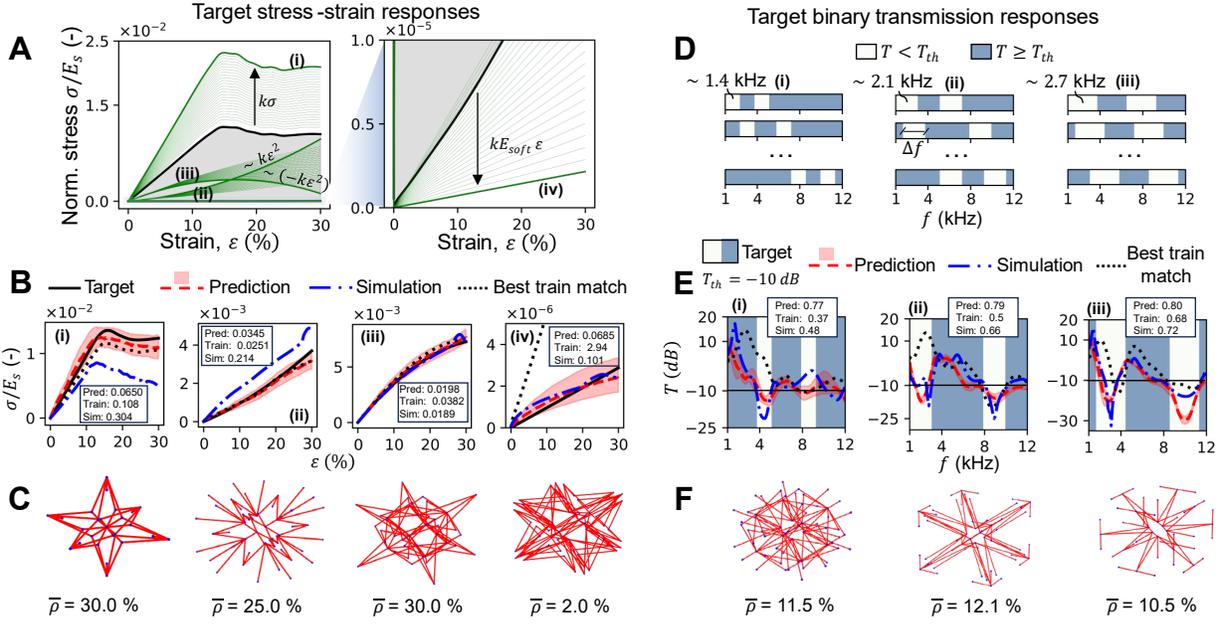

**Fig. 3. Design of structures with user-defined functional responses.** (**A**) Four types of unseen stress-strain curves. (i) Stiffer and stronger. (ii) Convex strain-hardening. (iii) Concave strain-hardening. (iv) Softer. (**B**) One example of inverse design for each curve type. (**C**) Corresponding generated graphs for the stress-strain responses in (B). (**D**) Target binary transmission sequences with two variable attenuation gaps of size $\Delta f = 1.4, 2.1, 2.7$ kHz, for type (i), (ii), (iii), respectively. (**E**) One example of inverse design for each sequence type. The transmission threshold $T_{th}$ is set to $-10\ dB$. (**F**) Corresponding generated graphs for the transmission responses in (E). The relative error and accuracy between the prediction and target ('Pred'), the best train match and target ('Train'), and between the simulation of the generated graph and target ('Sim') are reported for each example in the plots in (B) and (E).

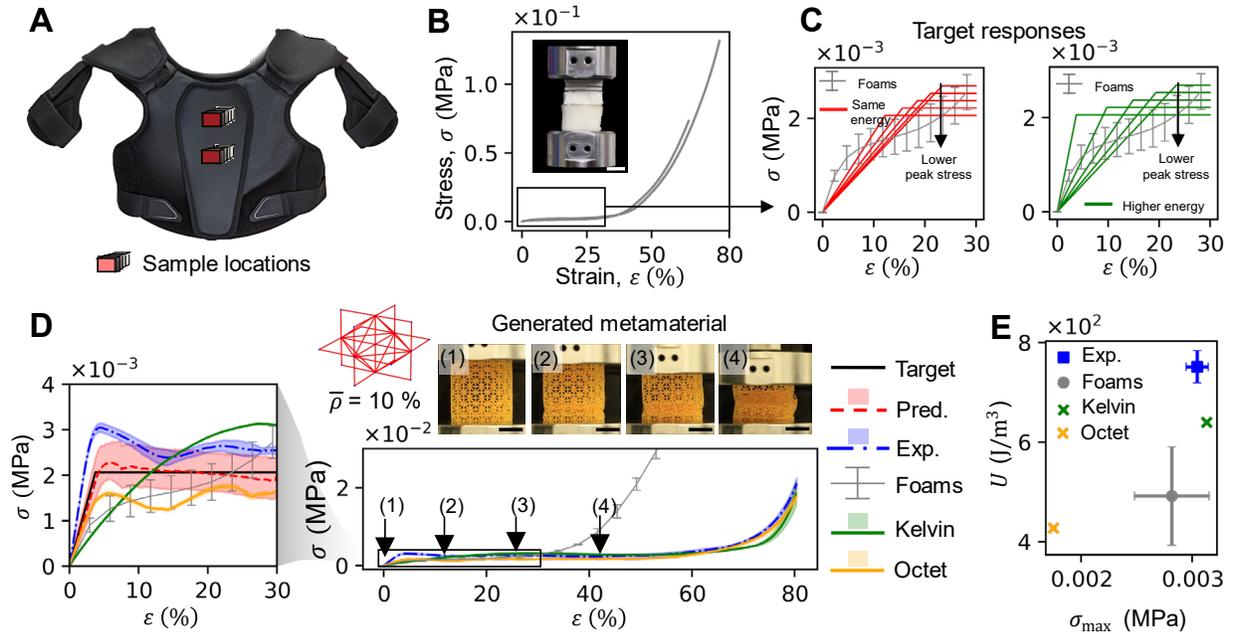

**Fig. 4. Design of energy-absorbing structures.** (**A**) Commercial lacrosse chest protector, showing the two locations where foam samples were extracted. (**B**) Nominal stress-strain curves of two representative four-layer foam sandwich samples, tested under quasi-static compression up to the onset of densification. The inset shows a snapshot of the sample during testing. Scale bar, 10 mm. (**C**) Target responses based on the foam sandwich baseline' stress-strain curve up to 30 % of strain. On the left plot, target curves for lower peak stress but same absorbed energy. On the right plot, target curves for lower peak stress and higher absorbed energy. The error bar identifies the deviation between tested foam samples. (**D**) Example of inverse-designed 5 × 5 × 5 metamaterial with target response corresponding to 30 % lower peak stress and 20 % higher energy absorption (with respect to the baseline). The main plot reports the experimentally measured stress-strain responses of the foam sandwich, the 3D-printed generated design and two classic 3D truss metamaterials, Kelvin foam and octet truss, up to densification. Four snapshots from a representative compressive test of the generated metamaterial are presented at different stages of deformation. Scale bar, 25 mm. The inset on the left shows a comparison of these curves along with the target and predicted responses up to 30% strain, the limit strain used for training GraphMetaMat. The labels 'Pred.' and 'Exp.' indicate the predicted and experimentally measured responses of the generated metamaterial. The Kelvin foam and octet-truss metamaterials were 3D-printed with same relative density $\bar{\rho} = 10\,\%$ as our design. (**E**) Energy absorption, $U$ vs. peak stress, $\sigma_{max}$ up to 30 % strain.

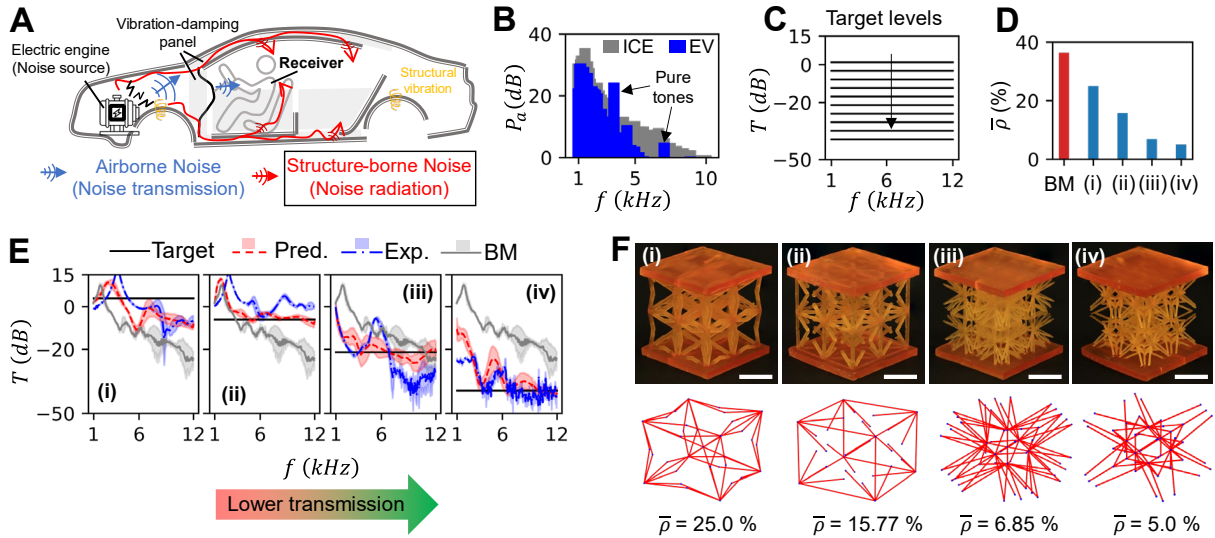

**Fig. 5. Design of noise-control structures.** (**A**) Sources of noise in electric vehicles. (**B**) Typical sound pressure level of internal combustion engines (ICE) and electrical vehicles (EV) in the frequency range 1 – 12 kHz, at 120 $km/h$ (*45*). The pure tonal peaks correspond to 4 and 8 times the number of poles in the electric motor. (**C**) Constant target transmission curves, from 5 to – 40 dB. (**D**) Relative density comparison between inverse-designed structures, (i) – (iv), and benchmark, BM (*46*). (**E**) Transmission responses obtained from the inverse design of noise-control structures with decreasing transmission target level ('Target'). Labels 'Pred.' and 'Exp.' indicate the predicted and experimentally measured responses of the generated metamaterial, respectively. The response of the benchmark, 'BM' is the same in the four plots. (**F**) 3D-printed inverse-designed structures with the corresponding graph and relative density, $\bar{\rho}$. Scale bar, 10 mm.

## Materials and Methods

Graph representation

Truss metamaterials are represented as graphs $G(V, E)$, i.e., a collection of nodes $V$ connected through edges $E$. The intersection between struts is represented in graph space by $V$, while the struts by $E$. We note that a dual representation of $G$ can also be used; however, we did not find any computational advantage and/or performance improvement in our case studies. The geometry of the structure is encoded into node $v_i \in V$ and edge $e_{ij} \in E$ features for each $i, j$ node. Any geometric features of the structure, such as node coordinates, struts' length, shape, orientation, and angle between struts, can be encoded into $v_i$ and $e_{ij}$. Based on empirical experiments on the model architecture used here (tables from S1 to S6), we observed that encoding the nodal coordinates as $v_i$ and the struts' length as $e_{ij}$ outperforms configurations with additional geometrical information. For the sake of simplicity, we assume circular struts with uniform radius. To uniquely determine the metamaterial' geometry, relative density $\bar{\rho}$ is the additional geometric parameter that has to be encoded into the graph representation. Unlike node coordinates or strut lengths, which take unique values for different nodes and edges, relative density is a global property for periodic metamaterials, describing all the strut radii with a single value. Hence, instead of encoding $\bar{\rho}$ as a node or edge feature, processed via the GNN, we input it into subsequent model's layers (see Supplementary Text "Forward model details"). The topology of the structure is inherently encoded by the graph connectivity.

Finite element simulations

The stress-strain and elastic wave transmission responses were collected via high-fidelity finite element (FE) simulations. To reduce the computational cost of training data collection, $2 \times 2 \times 2$ periodic metamaterials were simulated under quasi-static and dynamic vibration excitations (see Supplementary Text "Finite-size effect" for a sensitivity analysis on $M \times M \times M$ tessellations with $M > 2$).

*Stress-strain*

As schematically shown in Fig. 1D, to collect the stress-strain curves, we performed quasi-static compressive simulations of the metamaterials between two rigid plates. The structure is constrained to the bottom and top plate, while a vertical displacement $u^*$ is applied to the top plate, corresponding to the final macroscopic strain $\varepsilon_{max}$ via $u^* = \varepsilon_{max} ML$. $ML$ corresponds to the initial height of the structure, with $L$ being the unit cell size, and $M$ the number of unit cells. In our case, $M = 2$, and $L = 25$ mm. It is crucial to note that our inverse design framework is scale independent as far as the training data are consistent with the constitutive material's behavior. The stress is computed as $F/A$ where $F$ is the total reaction force measured at the bottom plate, and $A = ML \times ML$ is the cross-sectional area of the lattice. The applied strain is computed as $\varepsilon = u/(ML)$, where $u$ corresponds to the applied displacement on the top plate. As a post-processing phase, the collected responses are filtered to remove high-frequency numerical oscillations. The constitutive material is modeled as linear elastic with Young's modulus $E_s = 4.0$ $MPa$, and Poisson's ratio $\nu_s = 0.3$, with properties of the Formlabs Flexible 80A material. Owing to its large strain at failure, fracture of the constitutive material is not here considered (see Materials and Methods "Constitutive materials characterization"). Abaqus/Explicit is employed to simulate the structures up to 30 % of strain. Quasi-static conditions are ensured by limiting the kinetic energy within 1 % of the internal energy. Geometric nonlinearities and frictionless contact are modeled. To validate the explicit solver results, we compared the stress-strain responses and deformed shapes of a complex heterogeneous structure using an implicit solver (Abaqus/Standard) (fig. S10).

The structures are meshed using four-node tetrahedral elements (C3D4 in Abaqus) with average size $2R$, where $R$ is the beam's radius, depending on the structure and relative density. A mesh sensitivity analysis was performed to balance accuracy and speed. Strength and stiffness reported in the inset of Fig. 1D are defined as the peak stress and slope of the curve at $\varepsilon = 0.1$ %, respectively.

*Wave transmission*

The transmission curves were obtained via vibration simulations using the Solid Mechanics module in COMSOL Multiphysics 6.0. To resemble application conditions and to ensure experimental reproducibility, lattice-cored sandwich structures are considered, meaning the plates are not rigid but modeled with the same constitutive material (fig. S9). Here, we set unit cell size $L = 10$ mm and tessellation of unit cells $M = 2$. A sinusoidal excitation force is applied on the bottom plate within a 4.5 mm radius circular region, matching the shaker contact area in the experimental setup (see fig. S9B and Materials and Methods "Mechanical Testing"). A harmonic frequency sweep is performed in the range of 1–12 kHz. The wave transmission curve along the excitation direction, measured in decibel (dB), is then calculated by $20 \log_{10}(u_{out}/u_{in})$, where $u_{in}$ is the input displacement averaged over the shaker contact area and $u_{out}$ is the output displacement at the center point of the top plate. This output displacement corresponds to the displacement measured by the laser vibrometer (LDV) during experimental testing. The constitutive material, TMPTA, is modeled as viscoelastic with frequency-dependent storage Young's modulus $E'_s(f)$, and loss factor $\tan(\delta)\,(f)$, constant Poisson's ratio $\nu_s = 0.3$, and density $\rho_s = 1050$ kg/m$^3$. (see Materials and Methods "Constitutive materials characterization"). The structures are meshed using four-node tetrahedral elements with minimum and maximum size $R$ and $2R$, respectively, where $R$ is the beam's radius, depending on the structure and relative density. A mesh sensitivity analysis was performed to balance accuracy and speed. As a post-processing phase, in agreement with our measurement system (fig. S9B), noise floor was set to $-40$ dB, making the transmission responses flat for values below $-40$ dB.

Sample fabrication

Samples were fabricated by using a digital light 3D-printer Anycubic Photon Ultra and D2 (ANYCUBIC Technology Co., Ltd) for compressive and transmission response design, respectively. To reach large deformations without catastrophic failure, the samples for quasi-static compressive responses are made of a commercial photosensitive resin, Flexible 80A (Formlabs Inc., Somerville, MA). To improve printability, we added 0.0125wt% photoabsorber to it. With this resin, slice thickness and exposure time were set to 0.050 mm and 15 s, respectively. To reduce the viscoelastic damping effect at higher frequencies, the samples for vibration transmission responses are made of a brittle material, denoted as TMPTA. This latter is an in-house photosensitive resin composed of trimethylolpropane triacrylate (Sigma–Aldrich Inc., St. Louis, MO) with 0.0125wt% photoabsorber and 2wt% phenylbis(2,4,6-trimethylbenzoyl) phosphine oxide photoinitiator (Sigma–Aldrich Inc., St. Louis, MO). With this resin, slice thickness and exposure time are set to 0.040 mm and 5 s, respectively. To balance 3D-printing resolution, printing volume and testing machine maximum load capacity, the samples for stress-strain response are fabricated with a unit cell size of 25 mm, being scale independent at mm-scale. On the contrary, the samples for wave transmission response design are fabricated with an as-designed unit cell size of 10 mm. After fabrication, all samples are cleaned with ethanol and dried in a dark environment for at least 24 hours.

Constitutive materials characterization

For quasi-static properties, dog-bone samples made of flexible resin (Flexible 80A) were fabricated employing the same printer and printing parameters used for the lattice samples. The samples were tested under quasi-static uniaxial tensile loading using a universal testing machine, Instron 5944 (Instron Corporation, Norwood, MA). The strain rate was set to $10^{-3} s^{-1}$. Fig. S8B shows the resulting stress-strain curves for three different printing directions. Although a certain degree of anisotropy is found for strength and strain at failure, we assume our structures will not fail up to 30 % of macroscopic strain (based on experimental observations, see Fig. 4D). Accordingly, a linear elastic material model with $E_s = 4$ MPa was fitted on these data.

Dynamic mechanical analysis (DMA, Q800 model, TA Instruments; fig. S8D) was used to characterize the viscoelastic properties of the material TMPTA, including the storage modulus $E'_s$ and the loss modulus ($E''_s$). The loss factor, $\tan(\delta)$, which is the ratio between $E''_s$ and $E'_s$, was also obtained. Specimens with dimensions of 25 mm length, 8 mm width, and 1 mm thickness were used for DMA measurements. A dynamic displacement with an oscillation amplitude of 5 μm was applied to the specimens over a frequency range of 0.1 Hz to 100 Hz, with 16 points sampled on a logarithmic scale. The measurements were conducted at temperatures ranging from -2°C to 22°C, with intervals of 3°C. The time-temperature superposition (TTS) principle was used to estimate the master curve by shifting each measured isotherms along the frequency axis to align with the selected reference temperature of 22°C (*51*). Master curves for $E''_s$, $E'_s$ and the corresponding $\tan(\delta)$ are obtained from 0.1 Hz to 10 kHz (fig. S8, E to G) and are next used for modeling the transmission curve in COMSOL. At least three samples per constitutive material were tested.

Mechanical testing

All quasi-static compression tests were performed by using a universal testing machine, Instron 5944 (Instron Corporation, Norwood, MA). The samples are compressed between the stationary and moving steel plates. To resemble the boundary conditions used in the training dataset, the samples are fixed on the two plates. Force-displacement curves are measured by the Instron load cell with a maximum capacity of 2000 N and the built-in crosshead encoder. Stress-strain curves are computed analogously to what is done for FE simulations. The strain rate for all tests is set to $10^{-3} s^{-1}$.

Vibration tests were performed by using an electrodynamical shaker, LDS (Ling Dynamic Systems, Ltd, UK), and a laser scanning vibrometer Polytec PSV-500 (Polytec Inc.). The sample was fixed to the shaker using a clear epoxy (Devcon). As schematically reported in fig. S9B, to vertically excite the samples, a sine sweep voltage, from 500 to 12500 Hz, is applied to the shaker. The following parameters were used: total sweep time of 65.54 s, amplitude voltage 5 V, FFT measurement mode with average of the magnitude on three measurements, sampling frequency 31.25 kHz, and frequency resolution ~15 mHz. The laser vibrometer is used to measure the acceleration spectrum $A(f)$ of a vibrating point. We first measure the acceleration of the shaker $A_{\text{shaker}}(f)$ without any sample. Then, to resemble the boundary conditions used in the training dataset, we measure the acceleration of the center point of the top plate of the sample $A_{\text{structure}}(f)$ (fig. S9A). From the definition of wave transmission, we finally compute the transmission response as $T(f) = A_{\text{structure}}(f)/A_{\text{shaker}}(f)$, where $A_{\text{shaker}}(f)$ represents the acceleration input to structure's bottom plate. The noise floor of the system is around $-40$ dB. For quasi-static compression and vibration tests, at least three samples per structure were tested.

GraphMetaMat

The proposed inverse-design framework is implemented in Python 3.9, within the PyTorch environment. Details are reported in the Supplementary Text.

Target responses
*Stress-strain*

The user-defined target curves reported in Fig. 3A are designed to provide enough diversity of out-of-distribution responses whose corresponding structure is not known a priori. Curve type (i) is constructed by rescaling the response of the stiffest and strongest lattice in the dataset by a factor $k \in [1.1, 2]$. Such response resembles an elastic perfect-plastic behavior. Type (ii) are convex stiffening curves following $\sigma = 0.1 E_{stiff}(\varepsilon + k\varepsilon^2)$, with $k \in [1, 10]$. Type (iii) are concave curves obtained by $\sigma = 0.5 E_{stiff}(\varepsilon - k\varepsilon^2)$, where $E_{stiff}$ is the stiffness of the most rigid structure in the dataset, and $k \in [1, 3]$. Curve type (iv) is a linear response obtained by $\sigma = k E_{soft} \varepsilon$, where $E_{soft}$ is the stiffness of the most compliant lattice in the dataset, $\varepsilon$ is the strain, and $k \in [0.1, 0.9]$. 40,000 curves were generated for training and testing the inverse model by linearly sampling $k$ with 10,000 points for each curve type.

The application-oriented target curves shown in Fig. 4C answer the need for higher energy absorption and lower peak stress by resembling an elastic perfect-plastic behavior. Using the quasi-static compressive response of foams in commercial chest protectors as baseline, 5,000 curves were generated by combining peak stress reduction in the range $5 - 30$ % and energy absorption increase in the range $0 - 20$ %. Given an energy absorption gain, i.e., larger area underneath the curve, the stiffness of the structure is accordingly adjusted for each peak stress reduction.

*Wave transmission*

The target binary sequences shown in Fig. 3D are constructed to inverse design metamaterials with tunable attenuation gaps, i.e., low transmission $T(f)$ in specific frequency ranges. The vector sequences are generated by first discretizing the frequency range into 16 intervals, and initializing the vectors with '1's. Then, two attenuation gaps, i.e., parts of the vectors filled with '0's, of size $0 < n < 8$, are randomly dispersed into the sequences without overlapping. The gaps have a frequency size $\Delta f = \frac{n}{16} * 11$ kHz. By varying $n$, we construct three types of sequences (i), (ii), and (iii), with $\Delta f \sim 1.4, 2.1, 2.7$ kHz, respectively. 55, 28, and 10 sequences were generated for type (i), (ii), and (iii), respectively.

The application-oriented target curves in Fig. 4H are constant transmission responses from 5 down to – 40 dB, in the frequency range $1 - 12$ kHz. In total, we generated 1,000 curves for training and testing our framework. Rather than designing metamaterials with constant transmission, with these target curves, we aim at challenging GraphMetaMat to design structures with tunable average transmission as well as to progressively discover structures with low transmission.